\documentclass[12pt]{article}
\usepackage{amsmath}
\usepackage{amssymb}

\input{tcilatex}

\begin{document}

\begin{center}
\textbf{DIRAC EQUATION IN FOUR TIME}

\smallskip \ 

\textbf{AND FOUR SPACE DIMENSIONS}

\smallskip \ 

\smallskip \ 

J. A. Nieto\footnote{%
niet@uas.edu.mx; janieto1@asu.edu} and M. Espinoza

\smallskip \ 

\textit{Facultad de Ciencias F\'{\i}sico-Matem\'{a}ticas de la Universidad
Aut\'{o}noma} \textit{de Sinaloa, 80010, Culiac\'{a}n Sinaloa, M\'{e}xico.}

\bigskip \ 

\bigskip \ 

\bigskip \ 

\textbf{Abstract}
\end{center}

The Dirac equation in four time and four space dimensions (or
(4+4)-dimensions) is considered. Step by step we show that such an equation
admits Majorana and Weyl solutions. In order to obtain the Majorana or Weyl
spinors we used a method based on the construction of Clifford algebra in
terms of 2x2-matrices. We argue that our approach can be useful in
supergravity, superstrings and qubit theory.

\bigskip \ 

\bigskip \ 

\bigskip \ 

\bigskip \ 

\bigskip \ 

\bigskip \ 

\bigskip \ 

\bigskip \ 


Pacs numbers: 04.60.-m, 04.65.+e, 11.15.-q, 11.30.Ly

September, 2016

\newpage

In this work, we explore a number of features of the Dirac equation [1] in
four time and four space dimensions (or $(4+4)$-dimensions). There are a
number of physical reasons to be interested in such a quest. First of all,
consider the splitting of the $(4+4)$-signature in the form $%
(4+4)=(3+1)+(1+3)$. One notes that $(3+1)$ is mere a changing signature of
our ordinary $(1+3)$-world. So, even from the beginning this splitting seems
to contain some kind of duality between the two signatures $(3+1)$ and $%
(1+3) $. Assuming that an electron "lives" in $(1+3)$-dimensions one wonders
one could be the corresponding dual electron in $(3+1)$-dimensions. At this
respect, it is worth mentioning that using a signature reversal a relation
between $(3+1)$ and $(1+3)$ signatures has already been investigated in the
context of string theory [2]-[3]. Secondly, in $(4+4)$-dimensions there
exist Majorana-Weyl spinors [4]-[7] (see also Ref. [8]) and therefore the $%
16 $ spinors complex components of the Dirac equation can be reduced to $4$
complex spinor components: the same number than an ordinary $\frac{1}{2}$%
-fermion in four dimensions. Analogue motivation may arise by considering
the splitting $(4+4)=(2+2)+(2+2)$. It has been shown that the $(2+2)$%
-signature is exceptional [9] and has interesting features such as
Majorana-Weyl spinors. Another source of physical interest emerges from the
fact that the $(4+4)$-dimensional theory may be obtained from dimensional
reduction to a $(5+5)$-dimensional theory which originates from the
so-called $M%
{\acute{}}%
$-theory [10]-[11] (see also Refs. [12]-[15]) which is defined in $(5+6)$%
-dimensions. In fact, upon spacelike compactification the $(5+6)$%
-dimensional theory leads to one Type II $A%
{\acute{}}%
$ and two Type II $B%
{\acute{}}%
$ string theories which "live" in $(5+5)$-dimensions [11]. Of course, in
this case, one must properly impose Majorana-Weyl constraints as in the case
of superstrings [16] and supergravity [17]. Moreover, it is interesting that
massless Dirac equation formulated in flat $(5+5)$ (or $(4+4)$) dimensions
may lead to massive spinors in $(1+3)$-dimensions [14]. In this context, it
has been shown that the triality automorphisms of $Spin(8)$ act on
Majorana-Weyl representations leading to relations among $%
(1+9)\leftrightarrow (5+5)\leftrightarrow (9+1)$ signatures, as well as
their corresponding transverse signatures $(0+8)\leftrightarrow
(4+4)\leftrightarrow (8+0)$ [13]. Finally, since the Dirac equation is a
hidden root of supersymmetry which, recently, in turn has been linked to
qubit theory [18] via the superqubits [19] (see also Refs. [20]-[25]) one
may expect that there must exist a description of the Dirac equation in
terms of qubit notion. In turn, $(4+4)$-dimensions have an interesting
connection with qubits and chirotopes [26]-[31]. So, we believe that
eventually the $(4+4)$-dimensional Dirac equation may shed some light on the
superqubit notion.

Here, we show that, in fact, the $(4+4)$-dimensional Dirac equation can be
linked to the qubit theory. Our strategy rests on the use of a basic basis
set of $2\times 2$-matrices elements of $M(2,R)$. We show that from this
structure one may obtain the Dirac gamma matrices and therefore the physical
states associated with the Dirac equation can be written in terms of qubit
sates.

Let us start considering the Dirac equation in any $(t+s)$-signature, namely

\begin{equation}
(\gamma ^{\hat{\mu}}\hat{p}_{\hat{\mu}}+m)\psi =0,  \label{1}
\end{equation}%
where $\gamma ^{\hat{\mu}}$ are the gamma matrices satisfying the Clifford
algebra

\begin{equation}
\gamma ^{\hat{\mu}}\gamma ^{\hat{\nu}}+\gamma ^{\hat{\nu}}\gamma ^{\hat{\mu}%
}=2\eta ^{\hat{\mu}\hat{\nu}}.  \label{2}
\end{equation}%
Here, $\eta ^{\hat{\mu}\hat{\nu}}$ is a $(t+s)$-dimensional flat diagonal
metric which depends on the signature $(t+s)$ ($t$ times and $s$ space
dimensions). For instance in $(1+3)$-dimensions, one has $\eta ^{\mu \nu
}=diag(-1,1,1,1)$, while in contrast in $(3+1)$-dimensions one gets $\eta
^{ab}=diag(-1,-1,-1,1)$. Note that (1) depends on the signature via the
expression (2).

In $(1+3)$-dimensions, the three more common representations of the gamma
matrices $\gamma ^{\mu }$ are the Weyl ($\gamma _{W}^{\mu }$), Dirac ($%
\gamma _{D}^{\mu }$) and Majorana ($\gamma _{M}^{\mu }$) representations.
Explicitly, considering the Pauli matrices

\begin{equation}
\begin{array}{ccccc}
\sigma _{1}\equiv \left( 
\begin{array}{cc}
0 & I \\ 
I & 0%
\end{array}%
\right) , &  & \sigma _{2}\equiv \left( 
\begin{array}{cc}
0 & -i \\ 
i & 0%
\end{array}%
\right) , &  & \sigma _{3}\equiv \left( 
\begin{array}{cc}
I & 0 \\ 
0 & -I%
\end{array}%
\right) ,%
\end{array}
\label{3}
\end{equation}%
one has

\begin{equation}
\begin{array}{ccc}
\gamma _{W}^{1}\equiv \left( 
\begin{array}{cc}
0 & 1 \\ 
1 & 0%
\end{array}%
\right) , &  & \gamma _{W}^{2}\equiv \left( 
\begin{array}{cc}
0 & \sigma ^{1} \\ 
-\sigma ^{1} & 0%
\end{array}%
\right) , \\ 
&  &  \\ 
\gamma _{W}^{3}\equiv \left( 
\begin{array}{cc}
0 & \sigma ^{2} \\ 
-\sigma ^{2} & 0%
\end{array}%
\right) , &  & \gamma _{W}^{4}\equiv \left( 
\begin{array}{cc}
0 & \sigma ^{3} \\ 
-\sigma ^{3} & 0%
\end{array}%
\right) .%
\end{array}
\label{4}
\end{equation}

\begin{equation}
\begin{array}{ccc}
\gamma _{D}^{1}\equiv \left( 
\begin{array}{cc}
1 & 0 \\ 
0 & -1%
\end{array}%
\right) , &  & \gamma _{D}^{2}\equiv \left( 
\begin{array}{cc}
0 & \sigma ^{1} \\ 
-\sigma ^{1} & 0%
\end{array}%
\right) , \\ 
&  &  \\ 
\gamma _{D}^{3}\equiv \left( 
\begin{array}{cc}
0 & \sigma ^{2} \\ 
-\sigma ^{2} & 0%
\end{array}%
\right) , &  & \gamma _{D}^{4}\equiv \left( 
\begin{array}{cc}
0 & \sigma ^{3} \\ 
-\sigma ^{3} & 0%
\end{array}%
\right) ,%
\end{array}
\label{5}
\end{equation}%
\begin{equation}
\begin{array}{ccc}
\gamma _{M}^{1}\equiv \left( 
\begin{array}{cc}
0 & -\sigma ^{2} \\ 
-\sigma ^{2} & 0%
\end{array}%
\right) , &  & \gamma _{M}^{2}\equiv \left( 
\begin{array}{cc}
i\sigma ^{3} & 0 \\ 
0 & i\sigma ^{3}%
\end{array}%
\right) , \\ 
&  &  \\ 
\gamma _{M}^{2}\equiv \left( 
\begin{array}{cc}
0 & \sigma ^{2} \\ 
-\sigma ^{2} & 0%
\end{array}%
\right) , &  & \gamma _{M}^{4}\equiv \left( 
\begin{array}{cc}
-i\sigma ^{1} & 0 \\ 
0 & -i\sigma ^{1}%
\end{array}%
\right) .%
\end{array}
\label{6}
\end{equation}%
By applying the unitary transformation%
\begin{equation}
V=\left( 
\begin{array}{cc}
1 & 1 \\ 
-1 & 1%
\end{array}%
\right) ,  \label{7}
\end{equation}%
to $\gamma _{W}^{\mu }$ one can obtain $\gamma _{D}^{\mu }$ from $\gamma
_{D}^{\mu }=V\gamma _{W}^{\mu }V^{-1}$. Correspondingly, one can get $\psi
_{D}=V\psi _{W}$. Similarly, by applying the unitary transformation

\begin{equation}
W=\left( 
\begin{array}{cc}
1 & \sigma ^{2} \\ 
-\sigma ^{2} & 1%
\end{array}%
\right) ,  \label{8}
\end{equation}%
one has $\gamma _{M}^{\mu }=W\gamma _{D}^{\mu }W^{-1}$ and $\psi _{M}=W\psi
_{D}$. \ Moreover, by combining these two results one can go from Weyl to
Majorana representation in the form $\gamma _{M}^{\mu }=WV\gamma _{W}^{\mu
}V^{-1}W^{-1}$ and $\psi _{M}=WV\psi _{W}$. Observe that the Pauli matrix $%
\sigma ^{2}$ plays an important role in these transformations.

When one changes the signature from $(1+3)$ to $(2+2)$ one notes that one
may simply change $\sigma ^{2}$ by $\rho ^{2}\equiv i\sigma ^{2}$ and
everything is modified accordingly. Of course, the difference is determined
by the flat metric in Clifford algebra (2) by changing $\eta _{\mu \nu
}=diag(-1,1,1,1)$ for $\eta _{\mu \nu }=diag(-1,1,-1,1)$. However something
peculiar happens when this change is performed. In the first case, the
physical state $\psi $ is complex but in the second case one may choose $%
\psi $ to be real. Further, in the first case, one may impose either
Majorana or Weyl conditions, but in the second case one finds that $\psi $
can be both Majorana and Weyl spinor.

In order to consider the Dirac equation (1) in higher dimensions let us
introduce the basic $2\times 2$-matrices,

\begin{equation}
\begin{array}{ccc}
\delta _{ij}\equiv \left( 
\begin{array}{cc}
1 & 0 \\ 
0 & 1%
\end{array}%
\right) , &  & \varepsilon _{ij}\equiv \left( 
\begin{array}{cc}
0 & 1 \\ 
-1 & 0%
\end{array}%
\right) , \\ 
&  &  \\ 
\eta _{ij}\equiv \left( 
\begin{array}{cc}
1 & 0 \\ 
0 & -1%
\end{array}%
\right) , &  & \lambda _{ij}\equiv \left( 
\begin{array}{cc}
0 & 1 \\ 
1 & 0%
\end{array}%
\right) .%
\end{array}
\label{9}
\end{equation}%
These matrices determine, in fact, a basis for any $2\times 2$-matrix
belonging to the set of $2\times 2$-matrices which we have denoted by $%
M(2,R) $. In fact, an arbitrary $2\times 2$-matrix can be written as

\begin{equation}
\Omega _{ij}=x\delta _{ij}+y\varepsilon _{ij}+r\eta _{ij}+s\lambda _{ij}.
\label{10}
\end{equation}%
A complex structure $\Omega _{ij}\longrightarrow z_{ij}=x\delta
_{ij}+y\varepsilon _{ij}$ is obtained from (10) by setting $r=0$ and $s=0$.
In fact, by setting $\delta _{ij}\rightarrow 1$ and $\varepsilon
_{ij}\rightarrow i$ one obtains a typical notation for a complex number,
namely $z=x+iy$. Note also that if $ad-bc\neq 0,$ that is if $\det \Omega
\neq 0$, then the matrices in $M(2,R)$ can be associated with the group $%
GL(2,R)$. If we further require $ad-bc=1$, then one gets the elements of the
subgroup $SL(2,R)$. It is worth mentioning that since one has the
isomorphisms $M(2,R)\sim C(2,0)\sim C(1,1)$ the fundamental matrices $\delta
_{ij},\eta _{ij},\lambda _{ij}$ and $\varepsilon _{ij}$ given in (9) no only
form a basis for $M(2,R)$ but also determine a basis for the Clifford
algebras $C(2,0)$ and $C(1,1)$. \ Moreover, one can show that $C(0,2)$ can
be constructed using the fundamental matrices (9) and the well known
Kronecker product $\otimes $. It turns out that $C(0,2)$ is isomorphic to
the quaternion algebra $H$. Indeed, there exist a theorem that establishes
that all the others higher dimensional Clifford algebras of arbitrary
signature $C(a,b)$ can be constructed from the building blocks $C(2,0),$ $%
C(1,1)$ and $C(0,2)$ (see Refs. [4]-[8]).

Using the Kronecker product $\otimes $ one can write the Pauli sigma
matrices (3) in the form

\begin{equation}
\begin{array}{ccccc}
\sigma _{10}=\lambda _{i_{2}j_{2}}\otimes \delta _{i_{1}j_{1}}, &  & \sigma
_{22}=\varepsilon _{i_{2}j_{2}}\otimes \varepsilon _{i_{1}j_{1}}, &  & 
\sigma _{30}=\eta _{i_{2}j_{2}}\otimes \delta _{i_{1}j_{1}},%
\end{array}
\label{11}
\end{equation}%
In this expression, the numbers $1,2,etc$ in the indices $%
i_{1},i_{2},..,i_{n}$ indicate the level $2^{n}$ of the corresponding index,
in other words, such numbers denote whether the matrices are $2\times
2,4\times 4,8\times 8,16\times 16,...etc$. For instance, the Pauli matrices
in (11) are, indeed, $4\times 4$-matrices.

One can now write a representation of the Dirac gamma matrices in terms of $%
\sigma _{10},\sigma _{22}$ and $\sigma _{30}$ as follows:%
\begin{equation}
\begin{array}{ccc}
\gamma _{300}=\eta _{i_{3}j_{3}}\otimes \delta _{i_{2}j_{2}}\otimes \delta
_{i_{1}j_{1}}, &  & \gamma _{210}=\varepsilon _{i_{3}j_{3}}\otimes \lambda
_{i_{2}j_{2}}\otimes \delta _{i_{1}j_{1}}, \\ 
&  &  \\ 
\gamma _{222}=\varepsilon _{i_{3}j_{3}}\otimes \varepsilon
_{i_{2}j_{2}}\otimes \varepsilon _{i_{1}j_{1}}, &  & \gamma
_{230}=\varepsilon _{i_{3}j_{3}}\otimes \eta _{i_{2}j_{2}}\otimes \delta
_{i_{1}j_{1}}.%
\end{array}
\label{12}
\end{equation}%
Up to a sign, a Majorana representation is obtained multiplying each
component in (12), on the left, by $(\gamma _{000}+\gamma _{222})$ and,
simultaneously on the right, by the inverse $\frac{1}{2}(\gamma
_{000}-\gamma _{222})$. In fact, one obtains

\begin{equation}
\begin{array}{ccc}
\gamma _{122}=\lambda _{i_{3}j_{3}}\otimes \varepsilon _{i_{2}j_{2}}\otimes
\varepsilon _{i_{1}j_{1}}, &  & \gamma _{032}=\delta _{i_{3}j_{3}}\otimes
\eta _{i_{2}j_{2}}\otimes \varepsilon _{i_{1}j_{1}}, \\ 
&  &  \\ 
\gamma _{222}=\varepsilon _{i_{3}j_{3}}\otimes \varepsilon
_{i_{2}j_{2}}\otimes \varepsilon _{i_{1}j_{1}}, &  & \gamma _{012}=\delta
_{i_{3}j_{3}}\otimes \lambda _{i_{2}j_{2}}\otimes \varepsilon _{i_{1}j_{1}}.%
\end{array}
\label{13}
\end{equation}%
Comparing (12) and (13) one notes that, in the first $2\times 2$-level, in
the Dirac representation there are a combined terms of $\delta _{i_{1}j_{1}}$
and $\varepsilon _{i_{1}j_{1}}$, while in the Majorana representation there
are only terms with $\varepsilon _{i_{1}j_{1}}$. In the traditional notation
of the gamma matrices this means that the Dirac representation is complex,
while the Majorana representation is pure imaginary (or pure real).

It turns out that the procedure can be generalized to any signature $d=t+s$.
For instance, in $(1+5)$-dimensions, (13) can be extended in the form

\begin{equation}
\begin{array}{ccc}
\gamma _{3000}=\eta _{i_{4}j_{4}}\otimes \delta _{i_{3}j_{3}}\otimes \delta
_{i_{2}j_{2}}\otimes \delta _{i_{1}j_{1}}, &  & \gamma _{2300}=\varepsilon
_{i_{4}j_{4}}\otimes \eta _{i_{3}j_{3}}\otimes \delta _{i_{2}j_{2}}\otimes
\delta _{i_{1}j_{1}}, \\ 
&  &  \\ 
\gamma _{2210}=\varepsilon _{i_{4}j_{4}}\otimes \varepsilon
_{i_{3}j_{3}}\otimes \lambda _{i_{2}j_{2}}\otimes \delta _{i_{1}j_{1}}, &  & 
\gamma _{2222}=\varepsilon _{i_{4}j_{4}}\otimes \varepsilon
_{i_{3}j_{3}}\otimes \varepsilon _{i_{2}j_{2}}\otimes \varepsilon
_{i_{1}j_{1}}, \\ 
&  &  \\ 
\gamma _{2230}=\varepsilon _{i_{4}j_{4}}\otimes \varepsilon
_{i_{3}j_{3}}\otimes \eta _{i_{2}j_{2}}\otimes \delta _{i_{1}j_{1}}, &  & 
\gamma _{2100}=\varepsilon _{i_{4}j_{4}}\otimes \lambda _{i_{3}j_{3}}\otimes
\delta _{i_{2}j_{2}}\otimes \delta _{i_{1}j_{1}}.%
\end{array}
\label{14}
\end{equation}%
While in $(3+5)$-dimensions the expressions (14) become

\begin{equation}
\begin{array}{c}
\gamma _{30000}=\eta _{i_{5}j_{5}}\otimes \delta _{i_{4}j_{4}}\otimes \delta
_{i_{3}j_{3}}\otimes \delta _{i_{2}j_{2}}\otimes \delta _{i_{1}j_{1}}, \\ 
\\ 
\gamma _{23000}=\varepsilon _{i_{5}j_{5}}\otimes \eta _{i_{4}j_{4}}\otimes
\delta _{i_{3}j_{3}}\otimes \delta _{i_{2}j_{2}}\otimes \delta _{i_{1}j_{1}},
\\ 
\\ 
\gamma _{22300}=\varepsilon _{i_{5}j_{5}}\otimes \varepsilon
_{i_{4}j_{4}}\otimes \eta _{i_{3}j_{3}}\otimes \delta _{i_{2}j_{2}}\otimes
\delta _{i_{1}j_{1}}, \\ 
\\ 
\gamma _{22210}=\varepsilon _{i_{5}j_{5}}\otimes \varepsilon
_{i_{4}j_{4}}\otimes \varepsilon _{i_{3}j_{3}}\otimes \lambda
_{i_{2}j_{2}}\otimes \delta _{i_{1}j_{1}}, \\ 
\\ 
\gamma _{22222}=\varepsilon _{i_{5}j_{5}}\otimes \varepsilon
_{i_{4}j_{4}}\otimes \varepsilon _{i_{3}j_{3}}\otimes \varepsilon
_{i_{2}j_{2}}\otimes \varepsilon _{i_{1}j_{1}}, \\ 
\\ 
\gamma _{22230}=\varepsilon _{i_{5}j_{5}}\otimes \varepsilon
_{i_{4}j_{4}}\otimes \varepsilon _{i_{3}j_{3}}\otimes \eta
_{i_{2}j_{2}}\otimes \delta _{i_{1}j_{1}}, \\ 
\\ 
\gamma _{22100}=\varepsilon _{i_{5}j_{5}}\otimes \varepsilon
_{i_{4}j_{4}}\otimes \lambda _{i_{3}j_{3}}\otimes \delta
_{i_{2}j_{2}}\otimes \delta _{i_{1}j_{1}}, \\ 
\\ 
\gamma _{21000}=\varepsilon _{i_{5}j_{5}}\otimes \lambda
_{i_{4}j_{4}}\otimes \delta _{i_{3}j_{3}}\otimes \delta _{i_{2}j_{2}}\otimes
\delta _{i_{1}j_{1}}.%
\end{array}
\label{15}
\end{equation}%
In this case the Majorana representation is obtained by multiplying, each
component in (15), in left and right, by $(\gamma _{00000}+\gamma _{22222})$
and $\frac{1}{2}(\gamma _{00000}-\gamma _{22222})$, respectively. One obtains

\begin{equation}
\begin{array}{c}
\gamma _{12222}=\lambda _{i_{5}j_{5}}\otimes \varepsilon
_{i_{4}j_{4}}\otimes \varepsilon _{i_{3}j_{3}}\otimes \varepsilon
_{i_{2}j_{2}}\otimes \varepsilon _{i_{1}j_{1}}, \\ 
\\ 
\gamma _{01222}=\delta _{i_{5}j_{5}}\otimes \lambda _{i_{4}j_{4}}\otimes
\varepsilon _{i_{3}j_{3}}\otimes \varepsilon _{i_{2}j_{2}}\otimes
\varepsilon _{i_{1}j_{1}}, \\ 
\\ 
\gamma _{00122}=\delta _{i_{5}j_{5}}\otimes \delta _{i_{4}j_{4}}\otimes
\lambda _{i_{3}j_{3}}\otimes \otimes \varepsilon _{i_{2}j_{2}}\otimes
\varepsilon _{i_{1}j_{1}}, \\ 
\\ 
\gamma _{00032}=\delta _{i_{5}j_{5}}\otimes \delta _{i_{4}j_{4}}\otimes
\delta _{i_{3}j_{3}}\otimes \eta _{i_{2}j_{2}}\otimes \varepsilon
_{i_{1}j_{1}}, \\ 
\\ 
\gamma _{22222}=\varepsilon _{i_{5}j_{5}}\otimes \varepsilon
_{i_{4}j_{4}}\otimes \varepsilon _{i_{3}j_{3}}\otimes \varepsilon
_{i_{2}j_{2}}\otimes \varepsilon _{i_{1}j_{1}}, \\ 
\\ 
\gamma _{00012}=\delta _{i_{5}j_{5}}\otimes \delta _{i_{4}j_{4}}\otimes
\delta _{i_{3}j_{3}}\otimes \lambda _{i_{2}j_{2}}\otimes \varepsilon
_{i_{1}j_{1}}, \\ 
\\ 
\gamma _{00322}=\delta _{i_{5}j_{5}}\otimes \delta _{i_{4}j_{4}}\otimes \eta
_{i_{3}j_{3}}\otimes \varepsilon _{i_{2}j_{2}}\otimes \varepsilon
_{i_{1}j_{1}}, \\ 
\\ 
\gamma _{03222}=\delta _{i_{5}j_{5}}\otimes \eta _{i_{4}j_{4}}\otimes
\varepsilon _{i_{3}j_{3}}\otimes \varepsilon _{i_{2}j_{2}}\otimes
\varepsilon _{i_{1}j_{1}}.%
\end{array}
\label{16}
\end{equation}%
Note that in this case once again the $2\times 2$-level involves only $%
\varepsilon _{i_{1}j_{1}}$ which means that the Majorana representation is
pure imaginary.

By convenience we wrote the gamma matrices without indices. But, in fact, in
four dimensions one can use the prescription

\begin{equation}
\begin{array}{ccc}
\gamma _{300}\rightarrow \gamma _{i_{1}i_{2}i_{3}j_{1}j_{2}j_{3}}^{(0)}, & 
& \gamma _{210}\rightarrow \gamma _{i_{1}i_{2}i_{3}j_{1}j_{2}j_{3}}^{(1)},
\\ 
&  &  \\ 
\gamma _{222}\rightarrow \gamma _{i_{1}i_{2}i_{3}j_{1}j_{2}j_{3}}^{(2)}, & 
& \gamma _{230}\rightarrow \gamma _{i_{1}i_{2}i_{3}j_{1}j_{2}j_{3}}^{(3)}.%
\end{array}
\label{17}
\end{equation}%
Hence, one finds that the Dirac equation (1) can be written as

\begin{equation}
(\gamma _{i_{1}i_{2}i_{3}j_{1}j_{2}j_{3}}^{(\mu )}\hat{p}_{\mu }+m_{0}\delta
_{i_{1}i_{2}i_{3}j_{1}j_{2}j_{3}})\psi ^{j_{1}j_{2}j_{3}}=0,  \label{18}
\end{equation}%
where $\delta _{i_{1}i_{2}i_{3}j_{1}j_{2}j_{3}}=\delta _{i_{3}j_{3}}\otimes
\delta _{i_{2}j_{2}}\otimes \delta _{i_{1}j_{1}}$. From this expression, it
is evident that one can generalize the procedure to any signature $d=t+s$ in
the form

\begin{equation}
(\gamma _{i_{1}i_{2}...i_{n}j_{1}j_{2}...j_{n}}^{(\hat{\mu})}\hat{p}_{\hat{%
\mu}}+m_{0}\delta _{i_{1}i_{2}...i_{n}j_{1}j_{2}...j_{n}})\psi
^{j_{1}j_{2}...j_{n}}=0,  \label{19}
\end{equation}%
where the indices $(\hat{\mu})$ run from $1$ to $d=t+s$ and $\delta
_{i_{1}...i_{n}j_{1}...j_{n}}=\delta _{i_{n}j_{n}}\otimes ...\otimes \delta
_{i_{1}j_{1}}$. One requires, of course, that the quantities $\gamma
_{i_{1}i_{2}...i_{n}j_{1}j_{2}...j_{n}}^{(\hat{\mu})}$ satisfy the Clifford
algebra, namely

\begin{equation}
\begin{array}{c}
\gamma _{i_{1}i_{2}...i_{n}}^{(\hat{\mu})~~~~~~~k_{1}k_{2}...k_{n}}\gamma
_{k_{1}k_{2}...k_{n}j_{1}j_{2}...j_{n}}^{(\hat{\nu})}+\gamma
_{i_{1}i_{2}...i_{n}}^{(\hat{\nu})~~~~~~k_{1}k_{2}...k_{n}}\gamma
_{k_{1}k_{2}...k_{n}j_{1}j_{2}...j_{n}}^{(\hat{\mu})} \\ 
\\ 
=2\eta ^{(\hat{\mu}\hat{\nu})}\delta _{i_{1}i_{2}...i_{n}j_{1}j_{2}...j_{n}}.%
\end{array}
\label{20}
\end{equation}%
Here, $\eta ^{(\hat{\mu}\hat{\nu})}=diag(-1-1,...,-1,+1,+1,...,+1)$ is a
flat metric in $(t+s)$-dimensions.

Let us introduce the basis

\begin{equation}
| j_{1}j_{2}...j_{n}>=| j_{1}>\otimes | j_{2}>\otimes ...\otimes | j_{n}>.
\label{21}
\end{equation}%
Thus, in general any physical state satisfying (19) can be written as (see
Refs. [18] and references therein)%
\begin{equation}
| \Psi >=\sum \limits_{j_{1},j_{2},...,j_{n}=0}^{1}\psi
^{j_{1}j_{2}...j_{n}}| j_{1}j_{2}...j_{n}>.  \label{22}
\end{equation}%
So, one has discovered that the spinors $\psi ^{j_{1}j_{2}...j_{n}}$ in the
Dirac equation (for any signature) may be considered as qubit states. For $3$%
-qubit one has

\begin{equation}
|\Psi > =\sum \limits_{j_{1},j_{2},j_{3}=0}^{1}\psi ^{j_{1}j_{2}j_{3}}|
j_{1}j_{2}j_{3}>,  \label{23}
\end{equation}%
while for $4$-qubit one finds%
\begin{equation}
|\Psi >=\sum \limits_{j_{1},j_{2},j_{3},j_{4}=0}^{1}\psi
^{j_{1}j_{2}j_{3}j_{4}}| j_{1}j_{2}j_{3}j_{4}>.  \label{24}
\end{equation}%
Comparing (18) with (23) one sees that the state associated with $\frac{1}{2}
$-spin particles (for instance, the electron state) in four dimensions can
be identified with a $3$-qubit state.

In order to obtain representation for the $(4+4)$-signature one may multiply
the term $\gamma _{22222}$ in (15) by $\gamma _{00002}$. One obtains the
representation

\begin{equation}
\begin{array}{c}
\gamma _{30000}=\eta _{i_{5}j_{5}}\otimes \delta _{i_{4}j_{4}}\otimes \delta
_{i_{3}j_{3}}\otimes \delta _{i_{2}j_{2}}\otimes \delta _{i_{1}j_{1}}, \\ 
\\ 
\gamma _{23000}=\varepsilon _{i_{5}j_{5}}\otimes \eta _{i_{4}j_{4}}\otimes
\delta _{i_{3}j_{3}}\otimes \delta _{i_{2}j_{2}}\otimes \delta _{i_{1}j_{1}},
\\ 
\\ 
\gamma _{22300}=\varepsilon _{i_{5}j_{5}}\otimes \varepsilon
_{i_{4}j_{4}}\otimes \eta _{i_{3}j_{3}}\otimes \delta _{i_{2}j_{2}}\otimes
\delta _{i_{1}j_{1}}, \\ 
\\ 
\gamma _{22210}=\varepsilon _{i_{5}j_{5}}\otimes \varepsilon
_{i_{4}j_{4}}\otimes \varepsilon _{i_{3}j_{3}}\otimes \lambda
_{i_{2}j_{2}}\otimes \delta _{i_{1}j_{1}}, \\ 
\\ 
\gamma _{22220}=\varepsilon _{i_{5}j_{5}}\otimes \varepsilon
_{i_{4}j_{4}}\otimes \varepsilon _{i_{3}j_{3}}\otimes \varepsilon
_{i_{2}j_{2}}\otimes \delta _{i_{1}j_{1}}, \\ 
\\ 
\gamma _{22230}=\varepsilon _{i_{5}j_{5}}\otimes \varepsilon
_{i_{4}j_{4}}\otimes \varepsilon _{i_{3}j_{3}}\otimes \eta
_{i_{2}j_{2}}\otimes \delta _{i_{1}j_{1}}, \\ 
\\ 
\gamma _{22100}=\varepsilon _{i_{5}j_{5}}\otimes \varepsilon
_{i_{4}j_{4}}\otimes \lambda _{i_{3}j_{3}}\otimes \delta
_{i_{2}j_{2}}\otimes \delta _{i_{1}j_{1}}, \\ 
\\ 
\gamma _{21000}=\varepsilon _{i_{5}j_{5}}\otimes \lambda
_{i_{4}j_{4}}\otimes \delta _{i_{3}j_{3}}\otimes \delta _{i_{2}j_{2}}\otimes
\delta _{i_{1}j_{1}}.%
\end{array}
\label{25}
\end{equation}%
One notes that in the first level of (25) all quantities contain $\delta
_{i_{1}j_{1}}$. So, in fact, $\delta _{i_{1}j_{1}}$ is mere a amplification
of (25) and therefore can be dropped leaving

\begin{equation}
\begin{array}{c}
\gamma _{3000}=\eta _{i_{5}j_{5}}\otimes \delta _{i_{4}j_{4}}\otimes \delta
_{i_{3}j_{3}}\otimes \delta _{i_{2}j_{2}}, \\ 
\\ 
\gamma _{2300}=\varepsilon _{i_{5}j_{5}}\otimes \eta _{i_{4}j_{4}}\otimes
\delta _{i_{3}j_{3}}\otimes \delta _{i_{2}j_{2}}, \\ 
\\ 
\gamma _{2230}=\varepsilon _{i_{5}j_{5}}\otimes \varepsilon
_{i_{4}j_{4}}\otimes \eta _{i_{3}j_{3}}\otimes \delta _{i_{2}j_{2}}, \\ 
\\ 
\gamma _{2221}=\varepsilon _{i_{5}j_{5}}\otimes \varepsilon
_{i_{4}j_{4}}\otimes \varepsilon _{i_{3}j_{3}}\otimes \lambda _{i_{2}j_{2}},
\\ 
\\ 
\gamma _{2222}=\varepsilon _{i_{5}j_{5}}\otimes \varepsilon
_{i_{4}j_{4}}\otimes \varepsilon _{i_{3}j_{3}}\otimes \varepsilon
_{i_{2}j_{2}}, \\ 
\\ 
\gamma _{2223}=\varepsilon _{i_{5}j_{5}}\otimes \varepsilon
_{i_{4}j_{4}}\otimes \varepsilon _{i_{3}j_{3}}\otimes \eta _{i_{2}j_{2}}, \\ 
\\ 
\gamma _{2210}=\varepsilon _{i_{5}j_{5}}\otimes \varepsilon
_{i_{4}j_{4}}\otimes \lambda _{i_{3}j_{3}}\otimes \delta _{i_{2}j_{2}}, \\ 
\\ 
\gamma _{2100}=\varepsilon _{i_{5}j_{5}}\otimes \lambda _{i_{4}j_{4}}\otimes
\delta _{i_{3}j_{3}}\otimes \delta _{i_{2}j_{2}}.%
\end{array}
\label{26}
\end{equation}%
In turn this means that the physical states $\psi
^{j_{1}j_{2}.j_{3}j_{4}j_{5}}$ can be reduced to $\psi
^{j_{2}j_{3}.j_{4}j_{5}}$. In the usual notation, this is equivalent to
reduce the complex spinor to real spinor. Moreover, observe that the
matrices in (26) are now $16\times 16$ instead of the $32\times 32$ as in
(15). Correspondingly $\psi ^{j_{2}j_{3}.j_{4}j_{5}}$ has now only $16$
components. This means that if in addition one imposes the Weyl condition on 
$\psi ^{j_{2}j_{3}.j_{4}j_{5}}$ then one obtains only $8$ real components;
surprisingly the same number of components of the Dirac spinor in $(1+3)$%
-dimensions. So one wonders, as in in Ref. [14], whether a massless
Majorana-Weyl fermion in $(4+4)$-dimensions corresponds to a massive fermion
in $(1+3)$-dimensions.

One can further clarify our approach by the following arguments. Suppose one
starts with the two matrices

\begin{equation}
\begin{array}{c}
\gamma _{1}=\lambda , \\ 
\gamma _{2}=\varepsilon .%
\end{array}
\label{27}
\end{equation}%
(By convenience we shall not write the indices in the matrices $\delta
,\varepsilon ,$ $\eta $ and $\lambda $ given in (9).) These two matrices can
be associated with the $(1+1)$-signature because $(\gamma _{1})^{2}=\gamma
_{0}=\delta $ and $(\gamma _{2})^{2}=-\gamma _{0}=-\delta $ and $\gamma
_{1}\gamma _{2}+\gamma _{2}\gamma _{1}=0$. One can move to the $(1+2)$%
-signature using in addition to $\gamma _{1}$ and $\gamma _{2}$ the matrix $%
\gamma _{3}=\gamma _{2}\gamma _{1}$ in the form%
\begin{equation}
\begin{array}{c}
\gamma _{1}=\lambda , \\ 
\gamma _{2}=\varepsilon , \\ 
\gamma _{3}=\eta .%
\end{array}
\label{28}
\end{equation}%
Now, consider the extended structure%
\begin{equation}
\begin{array}{c}
\gamma _{10}=\lambda \otimes \delta , \\ 
\gamma _{21}=\varepsilon \otimes \lambda , \\ 
\gamma _{22}=\varepsilon \otimes \varepsilon , \\ 
\gamma _{23}=\varepsilon \otimes \eta ,%
\end{array}
\label{29}
\end{equation}%
obtained by introducing $\gamma _{10}$ in the first row and putting a $2$%
-index (or a $\varepsilon $) on the left hand side of each term in (28). One
can check that all matrices $\gamma $ in (29) satisfy the Clifford algebra
(2) with flat $(2+2)$-signature metric.

For the $(2+3)$-signature, one now considers the product $\gamma
_{30}=\gamma _{02}\gamma _{10}\gamma _{21}\gamma _{22}\gamma _{23}$ (Notes
that in this expression $\gamma _{02}$ acts as the imaginary unit $i$ in the
usual notation) and writes

\begin{equation}
\begin{array}{c}
\gamma _{10}=\lambda \otimes \delta , \\ 
\gamma _{21}=\varepsilon \otimes \lambda , \\ 
\gamma _{22}=\varepsilon \otimes \varepsilon , \\ 
\gamma _{23}=\varepsilon \otimes \eta , \\ 
\gamma _{30}=\eta \otimes \delta .%
\end{array}
\label{30}
\end{equation}
While for $(3+3)$-signature one has

\begin{equation}
\begin{array}{c}
\gamma _{100}=\lambda \otimes \delta \otimes \delta , \\ 
\gamma _{210}=\varepsilon \otimes \lambda \otimes \delta , \\ 
\gamma _{221}=\varepsilon \otimes \varepsilon \otimes \lambda , \\ 
\gamma _{222}=\varepsilon \otimes \varepsilon \otimes \varepsilon , \\ 
\gamma _{223}=\varepsilon \otimes \varepsilon \otimes \eta , \\ 
\gamma _{230}=\varepsilon \otimes \eta \otimes \delta .%
\end{array}
\label{31}
\end{equation}%
Following similar procedure, for $(3+4)$-signature one obtains

\begin{equation}
\begin{array}{c}
\gamma _{100}=\lambda \otimes \delta \otimes \delta , \\ 
\gamma _{210}=\varepsilon \otimes \lambda \otimes \delta , \\ 
\gamma _{221}=\varepsilon \otimes \varepsilon \otimes \lambda , \\ 
\gamma _{222}=\varepsilon \otimes \varepsilon \otimes \varepsilon , \\ 
\gamma _{223}=\varepsilon \otimes \varepsilon \otimes \eta , \\ 
\gamma _{230}=\varepsilon \otimes \eta \otimes \delta , \\ 
\gamma _{300}=\eta \otimes \delta \otimes \delta ,%
\end{array}
\label{32}
\end{equation}%
While for $(4+4)$-signature one gets%
\begin{equation}
\begin{array}{c}
\gamma _{1000}=\lambda \otimes \delta \otimes \delta \otimes \delta , \\ 
\gamma _{2100}=\varepsilon \otimes \lambda \otimes \delta \otimes \delta ,
\\ 
\gamma _{2210}=\varepsilon \otimes \varepsilon \otimes \lambda \otimes
\delta , \\ 
\gamma _{2221}=\varepsilon \otimes \varepsilon \otimes \varepsilon \otimes
\lambda , \\ 
\gamma _{2222}=\varepsilon \otimes \varepsilon \otimes \varepsilon \otimes
\varepsilon , \\ 
\gamma _{2223}=\varepsilon \otimes \varepsilon \otimes \varepsilon \otimes
\eta , \\ 
\gamma _{2230}=\varepsilon \otimes \varepsilon \otimes \eta \otimes \delta ,
\\ 
\gamma _{2300}=\varepsilon \otimes \eta \otimes \delta \otimes \delta .%
\end{array}
\label{33}
\end{equation}%
The method can be extended to $(n+n)$-dimensions. Simply, in the case of $%
(2n+(2n+1))$-signature one adds the gamma matrices $\gamma _{30...0}$ at the
end of the previous arrange of $(2n+2n)$-signature. While going from $%
(2n+(2n+1))$-signature to $((2n+1)+(2n+1))$ one adds in the first row a $%
\gamma _{10...0}$ and in all the other terms one inserts on the left level a 
$2$-index (or $\varepsilon $) of the previous signature. Observe that the
signature in all these cases is obtained by considering that if the number
of $\varepsilon $ is even the square of the specific gamma matrices is $%
+\delta $ and if the number of $\varepsilon $ is odd then the square of the
gamma matrices is $-\delta $. Here, one should remember that the quantities
are multiplied according to the level. Moreover, to calculate the complete
Clifford algebra (2) one notes that any two different gamma matrices commute
if the number of anticommuting terms is odd.

If instead of (27) one starts with

\begin{equation}
\begin{array}{c}
\gamma _{3}=\eta , \\ 
\gamma _{2}=\varepsilon ,%
\end{array}
\label{34}
\end{equation}%
and follow similar procedure inserting the gamma matrices $\gamma _{30...0}$
and $\gamma _{10...0}$ at the beginning and at the end respectively and in
all the other terms one inserts a $2$-index (or $\varepsilon $) in left
level of the previous signature one may prove that in $(4+4)$-signature one
obtains the result (26). In fact, in this case (26) corresponds to the Dirac
representation, while (33) refers to the Weyl representation. Note that in
each given $(n+n)$ \ dimension the $\gamma _{22...2}$ plays a key role, in a
sense it is the equilibrium gamma term between $\gamma _{10...0}$ and $%
\gamma _{30...0}$. Moreover, any other representation of the gamma matrices
satisfying (2) can be obtained from the Weyl, Dirac or Majorana
representation by similarity transformation.

One physical reason to be interested in these developments emerges from the
observation that massless fermions in $(4+4)$-dimension can lead to massive
one in $(1+3)$-dimension. In order to clarify this observation let us write
the massless Dirac equation (1) as

\begin{equation}
(\gamma ^{\mu }\hat{p}_{\mu }+\gamma ^{a}\hat{p}_{a})\psi =0.  \label{35}
\end{equation}%
Here, the terms $\gamma ^{\mu }\hat{p}_{\mu }$ and $\gamma ^{a}\hat{p}_{a}$
refer to $(1+3)$- and $(3+1)$-signature, respectively. Note that if $\gamma
^{a}\hat{p}_{a}$ determines a mass $m$ in the form

\begin{equation}
\gamma ^{a}\hat{p}_{a}\psi =m\psi ,  \label{36}
\end{equation}%
then (35) becomes the massive Dirac equation in $(1+3)$-dimensions and
therefore in the world of $(1+3)$-signature one has massive fermions. But
since (36) can also be written as

\begin{equation}
(\gamma ^{a}\hat{p}_{a}-m)\psi =0,  \label{37}
\end{equation}%
one discovers that the in the mirror $(3+1)$-world one also has massive
fermions, but with opposite signed mass. So, the mass $m$ is the quantity
linking the two scenarios with $(1+3)$ and $(3+1)$ signatures: In one case $%
m $ is positive and in the other is negative, respectively.

It is interesting that three index object $\eta _{ijk}$ with components 
\begin{equation}
\eta _{1ij}=\delta _{ij};\qquad \eta _{2ij}=\varepsilon _{ij}.  \label{38}
\end{equation}%
satisfies%
\begin{equation}
\eta _{ij1}=\eta _{ij};\qquad \eta _{ij2}=\lambda _{ij}.  \label{39}
\end{equation}%
In Refs [26] and [27], in analogy with the curved metric $g_{\mu \nu }=$ $%
e_{\mu }^{i}e_{\nu }^{j}\eta _{ij}$ , where $\eta _{ij}$ is a flat metric
and $e_{\mu }^{i}$ is a \textit{zweibeins, }the three-index curved metric
was proposed 
\begin{equation}
g_{\mu \nu \lambda }=e_{\mu }^{i}e_{\nu }^{j}e_{\lambda }^{k}\eta _{ijk}.
\label{40}
\end{equation}%
Since the matrices (38) and (39) are linked to Clifford algebras a
gravitational theory based on (40) may determine automatically a spin
structure, which is necessary for supersymmetric scenarios.

In turn, such a gravitational theory may be linked to oriented matroid
theory via the identifications

\begin{equation}
\begin{array}{ccc}
\delta _{ij}\rightarrow \{ \mathbf{V}^{1},\mathbf{V}^{2}\}, &  & \eta
_{ij}\rightarrow \{ \mathbf{V}^{1},\mathbf{V}^{3}\}, \\ 
&  &  \\ 
\lambda _{ij}\rightarrow \{ \mathbf{V}^{2},\mathbf{V}^{4}\}, &  & 
\varepsilon _{ij}\rightarrow \{ \mathbf{V}^{3},\mathbf{V}^{4}\}.%
\end{array}
\label{41}
\end{equation}%
where $\mathbf{V}^{1},\mathbf{V}^{2},\mathbf{V}^{3}$ and $\mathbf{V}^{4}$
are the columns of the matrix

\begin{equation}
V_{i}^{A}=\left( 
\begin{array}{cccc}
1 & 0 & 0 & 1 \\ 
0 & 1 & -1 & 0%
\end{array}%
\right) ,  \label{42}
\end{equation}%
with the index $A$ taking values in the set $E=\{1,2,3,4\}$. It turns out
that the subsets $\{ \mathbf{V}^{1},\mathbf{V}^{2}\}$, $\{ \mathbf{V}^{1},%
\mathbf{V}^{3}\}$, $\{ \mathbf{V}^{2},\mathbf{V}^{4}\}$ and $\{ \mathbf{V}%
^{3},\mathbf{V}^{4}\}$ are bases over the real of the matrix (42). One can
associate with these subsets the collection%
\begin{equation}
\mathcal{B}=\{ \{1,2\},\{1,3\},\{2,4\},\{3,4\} \},  \label{43}
\end{equation}%
which can be understood as a family of subsets of $E$. It is not difficult
to show that the pair $\mathcal{M}=(E,\mathcal{B})$ is a 2-rank self-dual
matroid (see Refs. [27] and [28] and references therein). The fact that we
can express $\mathcal{M}$ in the matrix form (42) means that this matroid is
representable (or realizable). Moreover, one can show that this matroid is
graphic and oriented. In the later case, the corresponding chirotope (see
[29] and references therein) is given by

\begin{equation}
\chi ^{AB}=\varepsilon ^{ij}V_{i}^{A}V_{j}^{B}.  \label{44}
\end{equation}%
Thus, we get, as nonvanishing elements of the chirotope $\chi ^{AB}$, the
combinations

\begin{equation}
\begin{array}{cccc}
12+, & 13-, & 24-, & 34+.%
\end{array}
\label{45}
\end{equation}%
The signs in (45) correspond to the determinants of the matrices $\delta
_{ij}$, $\eta _{ij}$, $\lambda _{ij}$ and $\varepsilon _{ij}$, which can be
calculated using (44). Therefore, what we have shown is that the basis of $%
M(2,R)$ as given in (9) admits an oriented matroid interpretation (see Refs.
[26]-[31] and references therein).

We believe that wherever one uses gamma matrices as supergravity and
superstrings our approach can be useful and therefore a link between these
physical scenarios with qubit theory can be established. In particular, a
search for a connection between the present developments and the
black-hole/qubit correspondence [20]-[25] may provide a source of motivation
for further investigation.

\bigskip \ 

\begin{center}
\textbf{Acknowledgments}
\end{center}

We would like to thank to C. Pereyra for helpful comments. This work was
partially supported by PROFAPI 2013.

\bigskip \

\end{document}